\documentclass[12pt,letterpaper]{article}

\usepackage[utf8]{inputenc}
\usepackage[english]{babel}
\usepackage{amsmath}
\usepackage{amsfonts}
\usepackage{amssymb}
\usepackage{graphicx}
\usepackage[left=1in,right=1in,top=1in,bottom=1in]{geometry}
\usepackage{changepage}
\usepackage{times}
\usepackage{cite}
\usepackage[colorlinks=true,citecolor=cyan,linkcolor=red,bookmarks=true,bookmarksnumbered=true]{hyperref}
\usepackage[font=footnotesize,labelfont=bf]{caption}
\usepackage{tikz}
\usetikzlibrary{arrows.meta}
\usepackage{subfigure}

\begin{document}
{
\flushleft\Huge\bf Supersymmetric Quantum Potentials Analogs of Classical Electrostatic Fields
}\vspace{5mm}

\begin{adjustwidth}{1in}{} 
{
\flushleft\large\bf Juan D Garc{\'i}a-Mu{\~n}oz$^1$\footnote{Author to whom correspondence should be addressed.} and A Raya$^{1,2}$ 
}
\vspace{2.5mm}
{
\par\noindent\small $^1$Instituto de F{\'i}sica y Matem{\'a}ticas, Universidad Michoacana de San Nicolás de Hidalgo, Edificio C-3, Ciudad Universitaria, Francisco J. M{\'u}jica S/N Col. Fel{\'i}citas del R{\'i}o, 58040 Morelia, Michoac{\'a}n, M{\'e}xico.\\
$^2$Centro de Ciencias Exactas, Universidad del Bío-Bío. Avda. Andrés Bello 720, Casilla 447, 3800708, Chillán, Chile.
}
\vspace{2.5mm}
{
\par\noindent\small Email: juan.domingo.garcia@umich.mx and alfredo.raya@umich.mx  
}
\vspace{5mm}
{
\par\noindent\bf Abstract
}
{ 
A relation between classical electrostatic fields and Shr{\"o}dinger-like Hamiltonians is evidenced. Hence, supersymmetric quantum potentials analogous to classical electrostatic fields can be constructed. Proposing an ansatz for the electrostatic potential as the natural logarithm of a nodeless function, it is demonstrated that the electrostatic fields fulfil the Bernoulli equation associated to a second-order confluent supersymmetric transformation. By using the so-called confluent algorithm, it is possible, given a charge density, to find the corresponding electrostatic field as well as the supersymmetric potentials. Furthermore, the associated charge density and the electrostatic field profile of Schr{\"o}dinger-like solvable potentials can be determined.    
}
\vspace{2mm}
\end{adjustwidth}

\section{Introduction} \label{S1}

Supersymmetric Quantum Mechanics (SUSY-QM) is an algebraic method allowing to intertwine two Schr{\"o}dinger-like Hamiltonians dubbed as SUSY partners~\cite{Gangopadhyaya2018,Junker2019,Bagchi2000,Cooper1995,Fernandez2010,Fernandez2019}. It has a direct connection with the factorization method~\cite{InfieldHull1951,Bogdan1984,Schrodinger1940,Schrodinger1941} and the Darboux transformation~\cite{Matveev1991}. By means of the so-called intertwining operator, which is in general a $k$th-order differential operator, one obtains the operational intertwining between the two SUSY partner Hamiltonians, thus defining a $k$th-order transformation~\cite{Nico2004}. In literature, the most common transformation is of the first-order, which has applications as generation of solvable quantum potentials~\cite{Diaz1999}, in particular shape-invariant potentials~\cite{Dutt1988}; determination of the generalized Heisenberg algebras for the SUSY partner potentials~\cite{Carballo2004,Hussin1999}, as well their coherent states~\cite{Veronique2007,D_az_Bautista_2019,D_az_Bautista_2020}; and more recently, in calculation of exact solutions for matrix Hamiltonians (analogous to the Dirac Hamiltonian) describing 2D materials, such as graphene~\cite{Hern_ndez_Ort_z_2011,Kuru2009,Concha_2018}; and the electron propagator in non-trivial magnetic backgrounds~\cite{Concha_S_nchez_2022}. Moreover, nowadays the second-order transformation has been proven  useful in the above quoted applications and more, including spectral manipulation~\cite{Contreras2008,Barnana2020}; and more specifically, in the description of the bilayer graphene~\cite{Juan2020,Juan2021}. 

On the other hand, from freshman courses in physics, electromagnetism is known to be a fundamental phenomenon for understanding the nature of our universe. However, the mathematical description of electric and magnetic fields, summarised in Maxwell equations, turns out be a hard nut to crack, generally speaking. A physical situation where it is more often possible to find the corresponding solutions is the static case. It is interesting that electrostatic systems as described by Maxwell equations lead to a first-order SUSY transformation fulfilling the standard superalgebra given by Witten~\cite{Witten1981}, in which the supersymmetric Hamiltonian is similar to the one-dimensional Dirac Hamiltonian~\cite{Gonzalez2019}. As a natural generalization, in this paper we show that electrostatic fields also develop a second-order supersymmetry transformation with a quadratic superalgebra. 

For a self-contained exposition of ideas, the organisation of the remaining of this work is as follows: A brief review of SUSY-QM will be given in Section \ref{SUSY}. In Section~\ref{S2} we describe the second-order transformation corresponding to electrostatic fields in a linear medium; Section~\ref{S3} shows some particular cases corresponding to known charge densities. These examples are used to illustrate the SUSY algorithm finding the associated intertwining Hamiltonians and the corresponding electrostatic fields. We further consider the quantum harmonic oscillator potential and obtain its corresponding charge density via supersymmetry. Finally, in Section~\ref{S4} we present our conclusions.   

\section{Supersymmetric Quantum Mechanics}\label{SUSY}
In Quantum Mechanics there exists an operational intertwining between two Schr{\"o}dinger-like Hamiltonians $H^{\pm}$ given by
\begin{equation} \label{E2.1}
    H^{\pm} = -\frac{d^{2}}{dx^{2}} + V^{\pm}(x),
\end{equation}
which satisfy the intertwining relation
\begin{equation}\label{E2.2}
    H^{+}L^{-} = L^{-}H^{-},
\end{equation}
with $L^{-}$ being a $k$th-order differential operator called intertwining operator. This SUSY transformation is described by means of the rules
\begin{equation}\label{E2.3}
    \left\{Q^{+},Q^{-}\right\}=H_{SS},\quad \left[Q^{\pm},H_{SS}\right]=0.
\end{equation}
It is standard to represent this algebra in terms of $2\times2$ matrices as follows
\begin{equation}\label{E2.4}
    Q^{+}=
    \begin{pmatrix}
    0 & L^{+} \\
    0 & 0 
    \end{pmatrix},\quad 
    Q^{-}=
    \begin{pmatrix}
        0 & 0 \\
        L^{-} & 0
    \end{pmatrix},\quad
    H_{SS}=
    \begin{pmatrix}
        L^{+}L^{-} & 0 \\
        0 & L^{-}L^{+}
    \end{pmatrix},
\end{equation}
where $Q^{\pm}$ are the so-called superchargers, $H_{SS}$ is the supersymmetric Hamiltonian and $L^{+} = (L^{-})^{\dagger}$. When the supersymmetric algebra in eq.~\eqref{E2.3} is satisfied, the Hamiltonians $H^{\pm}$ are called supersymmetric partners \cite{Nico2004} and it is obtained that
\begin{equation}\label{E2.5}
    \begin{aligned}
        &L^{+}L^{-} = (H^{+}-\varepsilon_{k})\dots(H^{+}-\varepsilon_{1}), \\
        &L^{-}L^{+} = (H^{-}-\varepsilon_{k})\dots(H^{-}-\varepsilon_{1}),
    \end{aligned}
\end{equation}
with $\varepsilon_{j}, j = 1,\dots,k$, being the factorization energies associated to the seed functions $u_{i}(x)$, eigenfunctions of the Hamiltonian $H^{-}$ with eigenvalue $\varepsilon_{i}$, which determine the SUSY transformation. If the Hamiltonians $H^{\pm}$ have eigenfunctions $\psi_{n}^{\pm}(x)$ associated to eigenvalues $E^{\pm}_{n}$, we have that
\begin{equation} \label{E2.6}
    \psi^{\pm}_{n}(x) = \sqrt{(E_{n}-\varepsilon_{k})\dots(E_{n}-\varepsilon_{1})}L^{\mp}\psi^{\mp}(x)
\end{equation}
It is important to note that this symmetry can take place for any two Hamiltonian operators as long as we can determine the corresponding intertwining operators $L^{\pm}$ in terms of the seed functions $u_{i}(x)$. In general, constructing a SUSY transformation is not a trivial task. However, low-order SUSY transformations can be developed. Below, we describe the corresponding first- and second-order SUSY algorithms.

\subsection{First-order SUSY-QM}

Typically, SUSY-QM is introduced by taking two Schr{\"o}dinger-like Hamiltonians of the form\cite{Gangopadhyaya2018,Junker2019}
{\begin{equation}\label{E2.7}
    H_{1}^{\pm} = -\frac{d^{2}}{dx^{2}} + V_{1}^{\pm}(x),
\end{equation}
while the intertwining operators $L^{\pm}_{1}$ are first-order differential operators given by
\begin{equation}\label{E2.8}
    L_{1}^{\pm} = \mp\frac{d}{dx} + \alpha(x), 
\end{equation}
where the function $\alpha(x)$ is referred to as the superpotential, which is written in terms of the seed function $u_{1}(x)$, a solution of the eigenvalue equation for $H^{-}$ associated to the factorization energy $\varepsilon_{1}$, i.e.,
\begin{equation}\label{E2.9}
    \alpha(x) = -\frac{u'_{1}(x)}{u_{1}(x)},\qquad H^{-}_{1}u_{1}(x) = \varepsilon_{1} u_{1}(x), 
\end{equation}
with $f'(x)\equiv df(x)/dx$. It is worth mentioning that the seed solution $u_{1}(x)$ must be a nodeless function inside the $x$-domain and the factorization energy $\varepsilon_{1} \leq E_{0}$, being $E_{0}$ the energy eigenvalue of the ground state of $H_{1}^{-}$. By using the intertwining relation in eq.~\eqref{E2.2}, it is possible to write the potentials $V^{\pm}_{1}$ in terms of the superpotential $\alpha(x)$ as follows
\begin{equation}\label{E2.10}
    V^{\pm}_{1} = \alpha^{2}(x) \pm \alpha'(x) +\varepsilon_{1}.
\end{equation}
Note that, in particular, the potential $V^{-}_{1}(x)$ and the superpotential $\alpha(x)$ satisfy the Riccati equation
\begin{equation} \label{E2.11}
    \alpha^{2}(x) - \alpha'(x) = V^{-}_{1}- \varepsilon_{1},
\end{equation}
which, using eq.~\eqref{E2.9}, leads to the eigenvalue equation for $H^{-}_{1}$.

On the other hand, the products of the intertwining operators $L^{\pm}_{1}$ turn out to be
\begin{equation} \label{E2.12}
    L_{1}^{-}L_{1}^{+} = H^{+}_{1}-\varepsilon_{1},\quad L_{1}^{+}L_{1}^{-} = H^{-}_{1}-\varepsilon_{1},
\end{equation}
and the eigenfunctions $\psi_{n}^{\pm}(x)$ are related in the form
\begin{equation} \label{E2.13}
    \psi^{\pm}_{n}(x) = \sqrt{E_{n}-\varepsilon_{1}}L^{\mp}_{1}\psi_{n}^{\mp}(x).
\end{equation}
Furthermore, the eigenfunctions $\psi^{\pm}_{\varepsilon_{1}}(x)$ of the Hamiltonians $H^{\pm}$, respectively, associated to the factorization energy $\varepsilon_{1}$ can be written as
\begin{equation}\label{E2.14}
    \psi^{+}_{\varepsilon_{1}}(x) = \frac{1}{u_{1}(x)},\quad \psi^{-}_{\varepsilon_{1}}(x) = u_{1}(x).
\end{equation} 
It is worth noting that, when one of the function $\psi^{\pm}_{\varepsilon_{1}}$ is square-integrable and the other one is not, the SUSY transformation is not isospectral; this case is know as unbroken SUSY (see Fig.~\ref{F1}). Conversely, when both functions are not square-integrable, the SUSY transformation is isospectral and we have the broken SUSY case. 

In general, because the products of the intertwining operators $L_{1}^{\pm}$ in eq.~\eqref{E2.12} are the Schr{\"o}dinger-like Hamiltonians $H_{1}^{\pm}$ (up to a constant), the operators $L_{1}^{\pm}$ also work as factorization operators of the Hamiltonians $H_{1}^{\pm}$, but no necessarily as ladder operators. Finally, for the first-order SUSY transformation the algebra defined in eq.~\eqref{E2.3} is know as superalgebra \cite{Witten1981}.

\subsection{Confluent second-order SUSY-QM}
 Confluent algorithm is a particular case of second-order Supersymmetric Quantum Mechanics where both factorization energies are equal, i.e., $\epsilon_{1} = \epsilon_{2}=\epsilon\in\mathbb{R}$. This is an algebraic method intertwining two Schr{\"o}dinger-like Hamiltonians $H_{2}^{\pm}$, whose form is analogous to the operators in eq.~\eqref{E2.7}. In this case, an intertwining relation, similar to the one written in eq.~\eqref{E2.2}, can be established. However, the intertwining operators $L^{\pm}_{2}$ are second-order differential operators. Specifically, they have the following form
\begin{equation} \label{E2.15}
    L^{-}_{2} = \frac{d^{2}}{dx^{2}} + \eta(x)\frac{d}{dx} + \gamma(x),\quad \gamma(x) = \left(\frac{\eta(x)}{2}\right)^{2} + \frac{\eta^{\prime}(x)}{2} + \left(\frac{\eta^{\prime}(x)}{2\eta(x)}\right)^{2} - \frac{\eta^{\prime\prime}(x)}{2\eta(x)}.
\end{equation}
From the previous equations, it can be observed that the confluent algorithm is defined by means of the function $\eta(x)$ \cite{Nico2004}  (see also \cite{Andrianov1993,Andrianov1995,Salinas2003,Salinas2005,Barnana2020,cs15a,cs15b,be16,cs17,sy18,bff12}). In this case, the said function fulfils the Bernulli equation
\begin{equation}\label{E2.16}
    \eta'(x) = \eta^{2}(x) + 2\beta(x)\eta(x),
\end{equation}
with $\beta(x) = u'(x)/u(x)$, being $u(x)$ the seed solution, fulfilling the stationary Schr{\"o}dinger-like equation for $H^{-}_{2}$, associated to the factorization energy $\varepsilon$. Solving the Bernulli equation \eqref{E2.16}, we have that
\begin{equation} \label{E2.17}
    \eta(x) = - \frac{w'(x)}{w(x)},\qquad w(x) = w_{0}(x) - \int\limits_{x_{0}}^{x}u^{2}(y)dy,
\end{equation}
where $x_{0}$ is a point in the appropriate $x$-domain and $w_{0}$ is a parameter that guarantees the function $w(x)$ remains nodeless. By substituting the expressions from eqs.~\eqref{E2.15} and \eqref{E2.16} in the corresponding intertwining relation, it is obtained that
\begin{equation} \label{E2.18}
        V^{-}_{2}(x) = \beta'(x) + \beta^{2}(x) +\varepsilon,\quad V^{+}_{2}(x) = V^{-}(x) + 2\eta^{\prime}(x).
\end{equation}

Furthermore, the products of the intertwining operators $L_{2}^{\pm}$ are
\begin{equation}\label{E2.19}
    L_{2}^{-}L_{2}^{+} = (H^{+}_{2}-\varepsilon)^{2},\quad L_{2}^{+}L_{2}^{-} = (H^{-}_{2}-\varepsilon)^{2},
\end{equation}
while the eigenfunctions $\psi^{\pm}_{n}(x)$ satisfy
\begin{equation} \label{E2.20}
    \psi^{\pm}_{n}(x) = |E_{n}-\varepsilon|L^{\mp}\psi^{\mp}(x)
\end{equation}
Furthermore, the eigenfunctions $\psi^{\pm}_{\varepsilon}(x)$ of the Hamiltonians $H^{\pm}_{2}$, respectively, corresponding to the factorization energy $\varepsilon$ can be written as 
\begin{equation} \label{E2.21}
    \psi^{+}_{\epsilon}(x) \propto \frac{u(x)}{w(x)}, \quad \psi^{-}_{\varepsilon}(x) = u(x).
\end{equation}
We must mention that it is a usual choice to select an the eigenfunction $\psi^{-}_{n}(x)$ of the Hamiltonian $H_{2}^{-}$ as seed solution. Thus, there are two kind of confluent SUSY transformations. When the eigenfunction $\psi^{+}_{\varepsilon}$ is square-integrable, the Hamiltonians $H_{2}^{\pm}$ have the same spectrum and the transformation is isospectral; while, when  $\psi^{+}_{\epsilon}$ is not square-integrable, the factorization energy does not belong to the spectrum of $H_{2}^{+}$, being the case of a confluent limit transformation. A consequence of this fact is that when the ground eigenfunction of a known Hamiltonian is chosen as seed solution, the confluent case is often confused with the first-order transformation (see Fig.~\ref{F1}), but the two algorithms are completely different from each other. Finally, from eq.~\eqref{E2.19}, we can observe the intertwining operators $L_{2}^{\pm}$ are  neither factorization nor ladder operators of the Hamiltonians $H_{2}^{\pm}$. For second-order SUSY transformation the algebra in  eq.~\eqref{E2.3} is so-called quadratic superalgebra.

\section{Supersymmetry of Electrostatic Fields} \label{S2}

In electrostatics, the electric field $\mathbf{E}$ obeys the Maxwell equations in an linear medium,
\begin{equation} \label{E1}
\begin{aligned}
\nabla\cdot\mathbf{D} = \rho, \\
\nabla\times\mathbf{E} = 0,
\end{aligned}
\end{equation}
where $\rho$ is the charge density and $\mathbf{D}$ is the electric displacement field given by
\begin{equation} \label{E2}
\mathbf{D} = \epsilon_{0}\mathbf{E} + \mathbf{P},
\end{equation}
with $\mathbf{P}$ being the electric polarization of the medium and $\epsilon_{0}$ the vacuum permittivity~\cite{Jackson2021}.

Let us consider an electrostatic field pointing out and changing only along a fixed direction, i.e., $\mathbf{E}(x) = E(x)\ \hat{x}$. Due to the Maxwell-Faraday eq.~\eqref{E1}, the electrostatic field $\mathbf{E}(x)$ must be the gradient of a scalar function $\varphi(x)$, called electrostatic potential. In other words,
\begin{equation} \label{E3}
\mathbf{E}(x) = - \nabla\varphi(x) \Rightarrow E(x) = -\frac{d\varphi(x)}{dx}.
\end{equation} 
Taking into account that such electrostatic potential $\varphi(x)$ can be expressed as the natural logarithm of a nodeless function $w(x)$, namely,
\begin{equation} \label{E4}
\varphi (x) = \varphi_{0}\ln \left[\frac{w(x)}{A}\right],
\end{equation}
with $\varphi_{0}$ and $A$ being constants with the appropriate units to ensure the correct dimensionality. Then, substituting the above expression into eq.~\eqref{E3}, it turns out that
\begin{equation} \label{E5}
E(x) = - \varphi_{0}\frac{w^{\prime}(x)}{w(x)}.
\end{equation} 
Thus, upon taking the derivative of the previous expression, we arrive at
\begin{equation}\label{E6}
E^{\prime}(x) = \frac{E^{2}(x)}{\varphi_{0}} - \varphi_{0}\frac{w^{\prime\prime}(x)}{w(x)}.
\end{equation}
It should be noted that, taking the function $w(x)$ as the seed solution, from eq.~\eqref{E5}, the quantity $E(x)/\varphi_{0}$ plays the role of the superpotential (see eq.~\eqref{E2.9}), and eq.~\eqref{E6} turns out to be a Riccati equation, as shown in eq.~~\eqref{E2.11}. Thus, defining a first-order SUSY transformation with a factorization energy $\varepsilon_{1}= 0$, we obtain exactly the case worked by González et al. \cite{Gonzalez2019}. Nevertheless, from eq.~\eqref{E5}, it is possible to construct a second-order SUSY transformation. Considering that the function $w(x)$ has the form given in eq.~\eqref{E2.17}, calculating its derivatives and substituting them back in eq.~\eqref{E6}, we obtain
\begin{equation} \label{E8}
\frac{E^{\prime}(x)}{\varphi_{0}} = \left(\frac{E(x)}{\varphi_{0}}\right)^{2} + 2\beta(x)\left(\frac{E(x)}{\varphi_{0}}\right).
\end{equation}
Equation~\eqref{E8} has the form of a Bernoulli equation \eqref{E2.16}, which describes a second-order confluent supersymmetric transformation where $E(x)/\varphi_{0}$ is taken as the function $\eta(x)$ in eq.~\eqref{E2.17}. 
In order to determine the confluent transformation defined in eq.~\eqref{E8}, notice that eq.~\eqref{E4} allows us to know $w(x)$ as a function of the electrostatic potential $\varphi(x)$ as
\begin{equation} \label{E13}
w(x) = Ae^{\varphi(x)/\varphi_{0}},
\end{equation}
whereas, from eq.~\eqref{E2.17}, it follows that $w'(x) = -u^{2}(x)$. Then, the seed solution can be expressed as 
\begin{equation} \label{E14}
u(x) = \sqrt{A\frac{E(x)}{\varphi_{0}}}e^{\varphi(x)/2\varphi_{0}} \propto e^{\int\beta(x) dx}.
\end{equation}
If we substitute $u(x)$ in the eigenvalue equation for $H_{2}^{-}$, we obtain that the corresponding factorization energy $\varepsilon = 0$. Note that the zero energy level is not necessarily the corresponding ground energy level of the spectrum of $H_{2}^{-}$. Moreover, given the form of $u(x)$ in eq.~\eqref{E14}, it is reasonable to think that the seed solution is asymptotically zero, since, away from the charge distributions, electrostatic potentials and fields tend to vanish. Thus, the seed solution should be square-integrable and then, the constant $A$ can be chosen so that the integral term in eq.~\eqref{E2.17} equals 1. Because the function $w(x)$ is nodeless, the constant $w_{0}$ does not lie in the range $(0,1)$. Then, comparing with eq.~\eqref{E13}, we get that $w_{0} = 0$. Consequently, the factorization energy level does not belong to the spectrum of the Hamiltonian $H^{+}_{2}$, i.e., we get a limit confluent SUSY transformation, while the corresponding function $\psi_{\varepsilon}^{+}(x) \propto u(x)/w(x)$ is not square-integrable. However, it could happen the seed solution in eq~\eqref{E14} is not square-integrable, though the eigenfunction $\psi_{\varepsilon}^{+}(x)$ is. Despite of this, since the integral term in eq.~\eqref{E2.17} must be convergent, it can also be concluded that $w_{0}=0$ in this case. Hence, the spectra of the Hamiltonians $H^{\pm}_{2}$ are the same except for the factorization energy level, see Fig.~\ref{F1}.
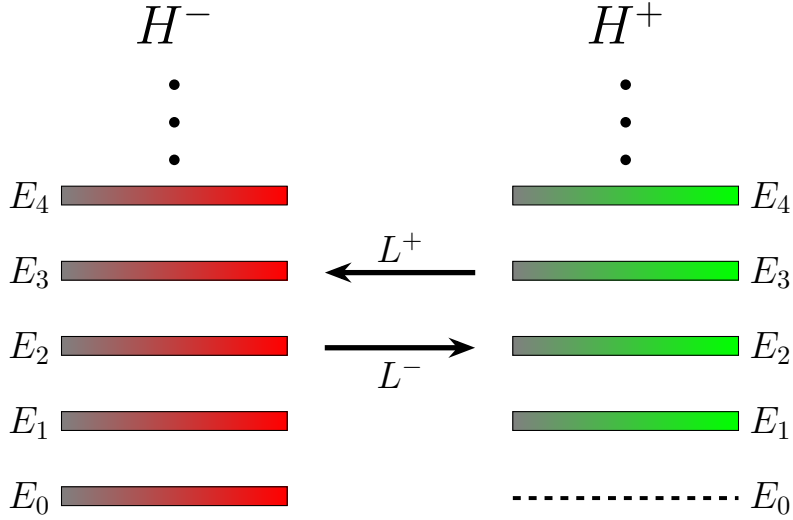
\begin{figure}[t]
\centering
\begin{tikzpicture}[step=1cm]

\foreach \x in {0,1}
{
\foreach \y in {0,1,2,3,4}
{
\ifnum \x=0 \shadedraw[right color=red] (6*\x,\y) rectangle (6*\x+3,\y+0.25); \draw (6*\x,\y+0.1) node[left]{\large $E_{\y}$};
\fi

\ifnum \x=1 
{
\ifnum \y=0 \draw[dashed,line width=1.5pt] (6*\x,\y+0.1)--(6*\x+3,\y+0.1); \draw (6*\x+3,\y+0.1) node[right]{\large $E_{\y}$};
\else \shadedraw[right color=green] (6*\x,\y) rectangle (6*\x+3,\y+0.25); \draw (6*\x+3,\y+0.1) node[right]{\large $E_{\y}$};
\fi
}
\fi

\ifnum \y<3 \fill (6*\x+1.5,4.6+0.5*\y) circle [radius=2pt];
\fi
}
}

\draw[-{Stealth[length=10pt]},line width=2pt] (3.5,2.1)--(5.5,2.1);
\draw[{Stealth[length=10pt]}-,line width=2pt] (3.5,3.1)--(5.5,3.1);

\draw (1.5,6) node[above]{\LARGE $H^{-}$} (7.5,6) node[above]{\LARGE $H^{+}$} (4.5,2.1) node[below]{\large $L^{-}$} (4.5,3.1) node[above]{\large $L^{+}$};
\end{tikzpicture}
\caption{Spectral scheme of the SUSY partner Hamiltonians $H^{\pm}$ and the intertwining linking them, when a second-order limit  confluent SUSY transformation is carried out, but also when a unbroken first-order SUSY transformation is developed. Both algorithms are constructed by using the ground state eigenfunction of $H^{-}$ as seed solution and, at first sight, they seem to be similar.} \label{F1}
\end{figure}   


Note that in the previous analysis, we have only worked the Maxwell-Faraday equation. However, the electric field $\mathbf{E}$ must also satisfy the Gauss law. Then,  Maxwell's equations~\eqref{E1} are transformed in the following system of equations
\begin{equation} \label{EX1}
    \eta'(x) = \frac{\rho(x)}{\epsilon\varphi_{0}},\quad \eta'(x) = \eta^{2}(x) + 2\beta(x)\eta(x),
\end{equation}
where $\eta(x)$ and $\beta(x)$ are the supersymmetric functions to be determined from the charge density $\rho(x)$. By substituting the first equation in the second one, we have that
\begin{equation} \label{EX2}
    \beta(x) = \frac{1}{2}\left(\frac{\rho(x)}{\int\rho dx} - \frac{1}{\epsilon\varphi_{0}}\int\rho dx\right).
\end{equation}
Thereby, in order to find $\eta(x)$, it is enough to use eq.~\eqref{EX2} in eq.~\eqref{EX1}. By returning to the original variables, it turns out that we have two solutions for the electric field, namely, 
\begin{equation} \label{EX3}
  E_{+}(x) = \frac{1}{\epsilon}\int\rho(x)dx,\quad E_{-}(x) = -\varphi_{0}\frac{\rho(x)}{\int\rho(x)dx}.
\end{equation}
Furthermore, the SUSY potential $V^{-}_{2}(x)$ can be directly calculated by means of eq.~\eqref{E2.18}. Thus, if we are given the charge density $\rho(x)$, by using the confluent algorithm, we obtain two SUSY partner potentials as well as the corresponding electrostatic field. It is worth mentioning $E_{\pm}(x)$ are mathematical solutions of the Maxwell equations~\eqref{E1} in an electrostatic situation. A complementary analysis is necessary to give a right interpretation of these solutions. One of the advantages of the SUSY-QM formalism is that it allows to approach in another different way the connection between electrostatic fields and quantum one-dimensional Hamiltonians. Specifically, assuming that we can obtain the eigenfunctions and energy eigenvalues of the Schr{\"o}dinger-like Hamiltonian $H^{-}_{2}$, we can take an eigenfunction as seed solution. Then, upon performing the confluent supersymmetric transformation, we can obtain the associated electrostatic field.
In the next section we develop some particular examples, which help  to illustrate the connection between electrostatic fields fulfilling  Maxwell eqs.~\eqref{E1} and the supersymmetric partner Schr{\"o}din{\-g}er-like Hamiltonians intertwined by means of the relation~\eqref{E2.2}. 

\section{Particular examples} \label{S3}

\subsection{Infinite charged sheet}
As a first example, let us consider a situation where the charge distribution is known and the electrostatic field is derived straightforwardly. We then look for the transformed quantum mechanical problem corresponding to a couple of SUSY partner potentials. We take a uniform surface charge distribution localized on an infinite sheet separating into two regions the space filled by a dielectric material with  permittivity constant $\epsilon$. In this case, $\rho(x) = \sigma\delta (x)$, $\sigma > 0$. Calculating the integral of this charge density and substituting in eq.~\eqref{EX2}, it turns out that 
\begin{equation} \label{E20}
\beta(x) = \frac{1}{2}\left(\delta(x) - \frac{\sigma}{\epsilon\varphi_{0}}\right).
\end{equation}
Consequently, using the previous form of $\beta(x)$ in eq.~\eqref{EX3}, the electrostatic fields  are given by
\begin{equation} \label{E21}
E_{+}(x) = \frac{\sigma}{\epsilon},\quad E_{-}(x) = -\varphi_{0}\delta(x).
\end{equation}
We associate the solution $E_{-}(x)$ with the field in the place where the charge lies, while the solution $E_{+}(x)$ is the field in the remaining of space. Evidently, it considers the flux contribution of the two infinite sheet surfaces, been half of it the correct magnitude of the electric field in each of the regions of space. In Fig.~\ref{F2}(a) we display a graph of the electrostatic field and the charge density. Substituting the function $\beta(x)$ in eq.~\eqref{E20} into eq.~\eqref{E2.18}, we arrive at the following supersymmetric partner potentials
\begin{equation} \label{E22}
V^{-}_{2}(x) = V^{+}_{2}(x) = \frac{\delta'(x)}{2} + \frac{1}{4}\left(\delta(x) - \frac{\sigma}{\epsilon\varphi_{0}}\right)^{2}.
\end{equation}
In Fig.~\ref{F2}(b) we sketch these potentials and $u(x)$, which can be obtained from eq.~\eqref{E14}, using eq.~\eqref{E20}. The properly normalized seed solution has the form
\begin{equation} \label{E23}
u(x) = 
\sqrt{\frac{\sigma}{2\epsilon\varphi_{0}}}e^{-\frac{\sigma}{2\epsilon\varphi_{0}}x + \frac{1}{2}},\quad x > 0.
\end{equation}
It is worth mentioning that, by construction, the confluent SUSY transformation guarantees the seed solution $u(x)$ is eigenfunction of $V^{-}_{2}(x)$ associated to the eigenvalue $\varepsilon = 0$. Moreover, since the infinite charged sheet in the electrostatic problem appears as an infinite potential barrier, eq.~\eqref{E22}, in the SUSY-QM problem, the $x$-domain splits into two regions and it is enough to analyze one of them, since the seed solution in the other region is analogous. The results obtained in this example are similar to those obtained in Ref.~\cite{Gonzalez2019}, where by means of first-order supersymmetric quantum mechanics, the case of an infinite charged sheet is also addressed. However, the confluent supersymmetric algorithm also allows us to find an extra solution $E_{-}(x)$ of the Maxwell equations, which is associated to the region where $E_{+}(x)$ is not physically appropriate. This solution cannot be obtained by means of the first-order SUSY QM.

\begin{figure}[t]
\centering
\subfigure[]{\includegraphics[height=8cm, width=8cm]{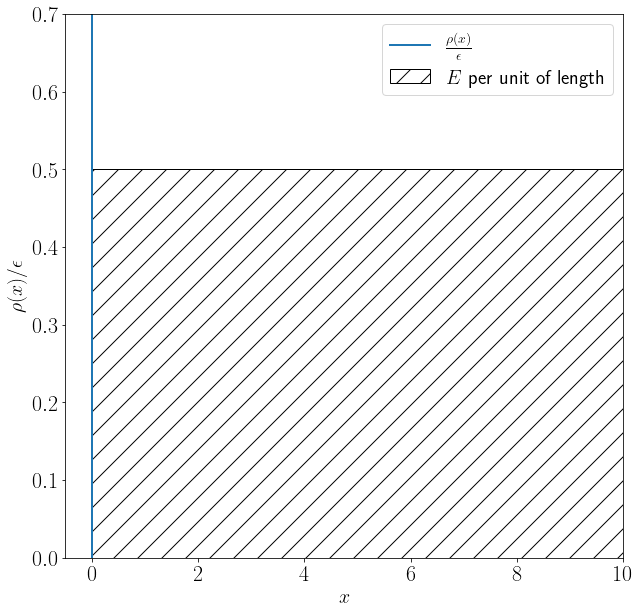}}
\subfigure[]{\includegraphics[height=8cm, width=8cm]{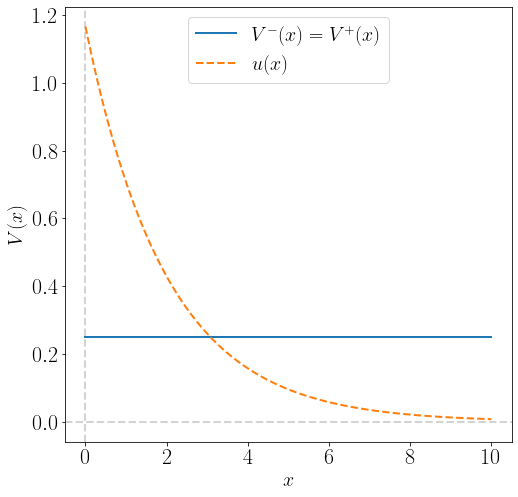}}
\caption{For the infinite charged sheet case: (a) A sketch of the charge density as well as the electrostatic field $E(x)$ per unit of length. (b) Plot of the SUSY partner potentials $V^{\pm}(x)$ and a representation of the seed solution $u(x)$ used to perform the confluent SUSY transformation. The scale of the graphs is fixed by the parameters $\epsilon = \varphi_{0} = \sigma = 1$.} \label{F2}
\end{figure} 

\subsection{Constant charge density}

As a second example, we consider a positive constant charge density $\rho = \rho_{0}$ uniformly distributed in an infinite dielectric box with finite width and a permittivity constant $\epsilon$. The function $\beta(x)$ is given by
\begin{equation} \label{E24}
\beta(x) = \frac{1}{2}\left(\frac{1}{x} - \frac{\rho_{0}}{\epsilon\varphi_{0}}\right).
\end{equation}
A straightforward calculation gives the electrostatic field solutions
\begin{equation} \label{E25}
E_{+}(x) = \frac{\rho_{0}}{\epsilon}x,\quad E_{-}(x) = -\frac{\varphi_{0}}{x}.
\end{equation} 
We can observe the solution $E_{+}(x)$ correspond to the electric field inside the dielectric box and the solution $E_{-}(x)$ is the electric field outside the box. In this case, $\varphi_{0}$ is a potential difference between a point $x\rightarrow \pm\infty$ and the corresponding surface of the box, which, since the electric field must be continuous,  equals  $\varphi_{0} = - \rho_{0}d^{2}/\epsilon$. The electric field and the charge density are displayed in Fig.~\ref{F3}(a). Performing the confluent transformation, we directly obtain the pair of SUSY partner potentials
\begin{equation} \label{E26}
\begin{aligned}
V_{2}^{-}(x) &= \omega^{2}x^{2} - \frac{1}{4x^{2}} - \omega, \\
V_{2}^{+}(x) &= \omega^{2}x^{2} - \frac{1}{4x^{2}} + \omega,
\end{aligned}
\end{equation}
with $\omega = (\rho_{0}/\epsilon\varphi_{0})< 0$. Figure~\ref{F3}(b) shows the SUSY partner potentials in eq.~\eqref{E26} as well as the seed solution, which can be written as
\begin{equation} \label{E27}
u(x) = \sqrt{\omega x}e^{-\frac{\omega}{4}x^{2}},\quad x>0.
\end{equation} 
We must mention that this seed solution is not square-integrable, while $\psi^{+}_{0}$ is. On the other hand, the potentials in eq.~\eqref{E26} have a ``centrifugal'' term and consequently, the appropriate $x$-domain is the range $[0,\infty)$. Analogous conclusions can be drawn for the range $(-\infty,0]$. Then, we can assume that the box lies in the range $[-d,d]$. Finally, $\psi_{\varepsilon}^{+}(x)$ is in agreement with the results obtained in Ref.~\cite{Sameer2011}, where the so-called isotonic potential $V(x) = \omega^{2}x^{2}/2 + g/2x^{2}$ is solved assuming $g$ constant. The potential $V^{+}_{2}(x)$ in eq.~\eqref{E26} is the limit case with $g=-1/4$, where there exists bound states. 

\begin{figure}[t]
\centering
\subfigure[]{\includegraphics[height=8cm, width=8cm]{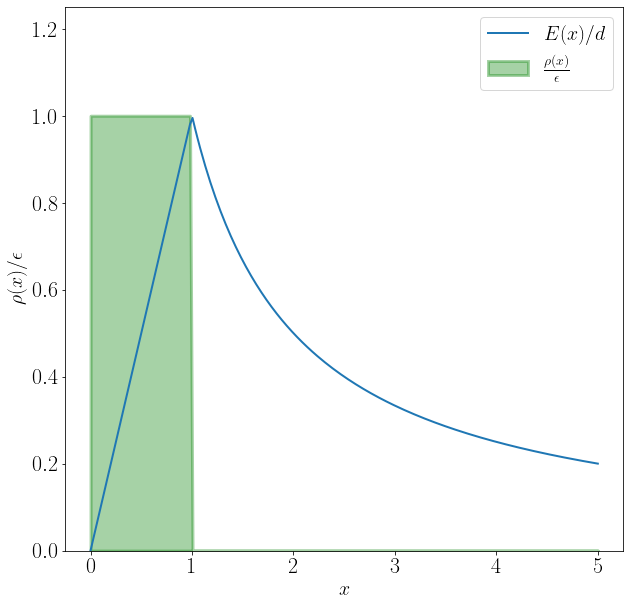}}
\subfigure[]{\includegraphics[height=8cm, width=8cm]{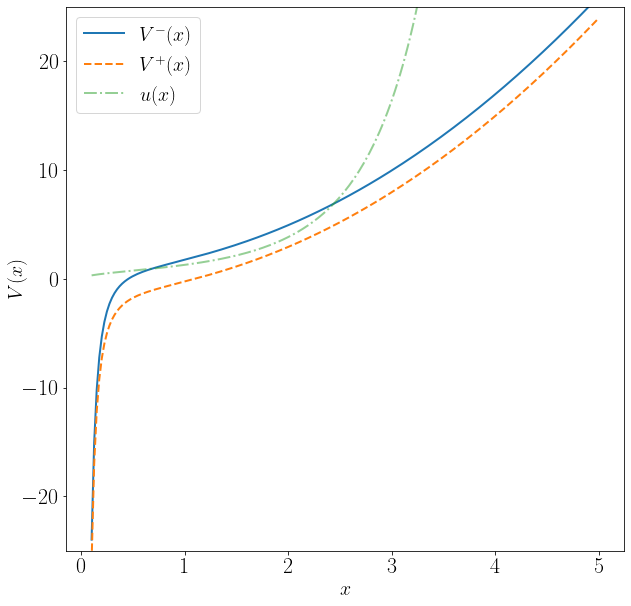}}
\caption{For the constant charge density: (a) Plot of the charge density divided by $\epsilon$ and the electrostatic field $E(x)/d$. (b) The SUSY partner potentials $V^{\pm}(x)$ as well a representation of the seed solution $u(x)$. The scale of the graphs is fixed by the parameters $\omega = \varphi_{0} = -1$ and $\epsilon = \rho_{0} = 1$.} \label{F3}
\end{figure} 

\subsection{Harmonic oscillator potential}

As a final example, we address the case where a quantum mechanical potential is known and look for the associated electrostatic field and the charge distribution originating it. For this purpose,
we choose the potential $V(x) = \omega^{2}x^{2}$,\ $\omega > 0$, $x\in \mathbb{R}$. Its well-known that the eigenfunctions $\psi_{n}^{-}(x)$ for this problem are given in terms of the Hermite polynomials, namely, 
\begin{equation} \label{E28}
\psi^{-}_{n}(x) = C_{n}e^{-\frac{\omega}{2}x^{2}}H_{n}(\sqrt{\omega}x),
\end{equation} 
with eigenvalues $E_{n}=\omega (2n+1)$. We can observe none of the bound states has a zero energy eigenvalue. Then, we need to subtract from the potential the energy of the bound state which will be used as seed solution. In other words, we consider a harmonic oscillator with an energy shift guaranteeing a zero energy bound state level. Taking the ground state eigenfunction as the seed solution and substituting in eq.~\eqref{E2.17}, the function $w(x)$ turns out to be
\begin{equation} \label{E29}
w(x) = -\frac{1}{2}\left(1 + \text{Erf}(\sqrt{\omega}x)\right),
\end{equation}
where $\text{Erf}(x)$ is the error function. Furthermore, using eq.~\eqref{E29} in eq.~\eqref{E2.17}, we obtain the function $\eta(x)$, and taking its derivative, the charge density has the following intricated form
\begin{equation} \label{E30}
\rho(x) = 4\epsilon\varphi_{0}\sqrt{\frac{\omega}{\pi}}\frac{e^{-\omega x^{2}}}{1 + \text{Erf}(\sqrt{\omega}x)}\left(\omega x +\sqrt{\frac{\omega}{\pi}}\frac{e^{-\omega x^{2}}}{1 + \text{Erf}(\sqrt{\omega}x)}\right),
\end{equation} 
Note that this density permeates the full space. Thus, we have an infinite charge distributed in it. Therefore, we should consider the electric field per unit of length rather than the field itself. Such linear electric field density behaves similarly to $\rho(x)/\epsilon$, see Fig.~\ref{F4}(a). Hence, this example reveals a highly non-trivial charge distribution which would be really tough to realize in the laboratory. However, from a theoretical viewpoint, it is interesting that by means of the supersymmetric transformation, it is associated to a simpler quantum mechanical problem. The SUSY partner potentials are given by 
\begin{equation} \label{E31}
V_{2}^{-}(x) = \omega^{2}x^{2} - \omega,\quad V^{+}_{2}(x) = V^{-}_2(x) + 2\frac{\rho(x)}{\epsilon\varphi_{0}.}
\end{equation} 
These potentials and the seed solution are shown in Fig.~\ref{F4}(b).

\begin{figure}[t]
\centering
\subfigure[]{\includegraphics[height=8cm, width=8cm]{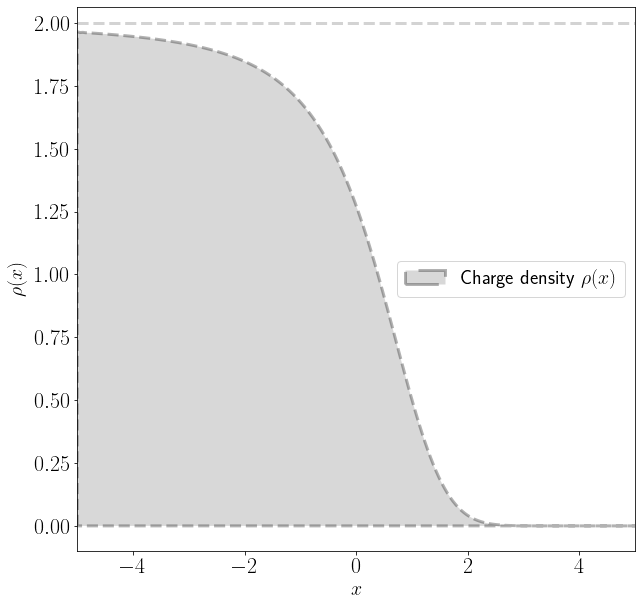}}
\subfigure[]{\includegraphics[height=8cm, width=8cm]{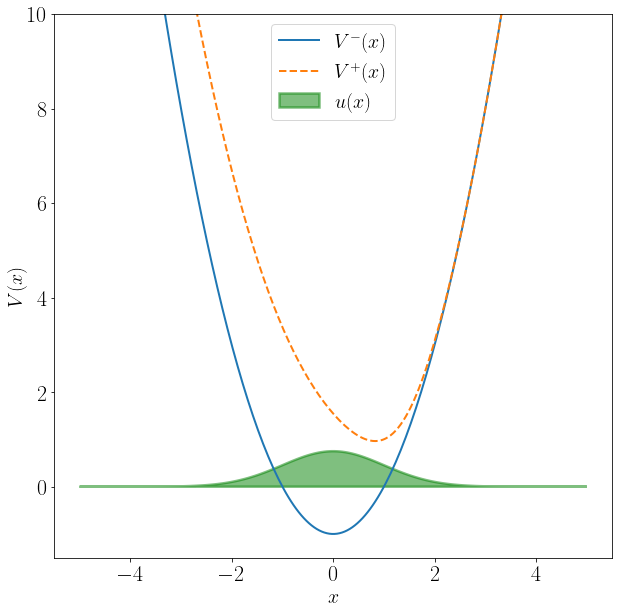}}
\caption{For the harmonic oscillator potential: (a) The corresponding charge density, which is asymptotically zero for $x\rightarrow \infty$, while for $x\rightarrow -\infty$ it tends to 2. (b) Plot of the SUSY partner potentials $V^{\pm}(x)$ and a representation of the seed solution $u(x)$ used to perform the confluent transformation. The scale of the graphs is set by the parameters $\omega = \varphi_{0} = \epsilon = 1$.} \label{F4}
\end{figure} 

\section{Conclusions} \label{S4}

The connection between electric fields fulfilling the electrostatic Maxwell equations and Schr{\"o}{\-}dinger-like second-order confluent supersymmetric partner Hamiltonians has been evidenced. Consequently,  supersymmetric quantum potentials analogs of classical electrostatic fields can be constructed. It is worth noticing that the confluent supersymmetric transformation defined in 
eq.~\eqref{E8} is not arbitrary, since it is carried out by means of an eigenfunction of $H^{-}_{2}$, associated to the zero eigenvalue, as seed solution and choosing the parameter $w_{0} = 0$. Moreover, the associated quadratic superalgebra in eq.~\eqref{E2.4} defines the matrix Hamiltonian $H_{SS}$, which is similar to the Hamiltonians describing 2D materials, in particular, bilayer graphene. It is important to observe that given the charge density, the confluent second-order SUSY-QM allows us to solve Maxwell equations and determine the electrostatic field in all the $x$-domain, a remarkable difference as compared with the case in which a first-order supersymmetric transformation is used, as can be seen in the first two examples of Section~\ref{S3}. Furthermore, for such particular profiles of charge density, the charge discontinuity is translated as infinite barrier or well in the quantum potential. On the other hand, by considering a solvable potential as $V^{-}_{2}(x)$, the associated charge density and the electric field can be obtained, as seen in the third example in previous Section, wherein, since there is an infinite charge, it is appropriate consider an linear electric field density rather than the field itself. We must mention that the charge density profile in eq.~\eqref{E30}, associated to the harmonic oscillator potential, could be pretty difficult to realize in the laboratory. Nevertheless, from a theoretical point of view, it is interesting to determine charge densities and electrostatic fields associated to solvable quantum potentials. Despite all the charge densities worked in this paper extend to infinity, the cases of an infinite charge sheet and an constant charge density correspond to  examples of  localized charge densities. Furthermore, one might, in any case, use conformal mappings of these extended distributions to finite regions and use the findings this article for that kind of problems. Also, one can consider, for example, a point charge density or a uniformly charged disk of finite radius but, since the algorithm developed here is one-dimensional, one would be careful to describe the system in the right coordinates and with the appropriate boundary conditions.

\section*{Acknowledgments}
We acknowledge financial support from CONACYT Project FORDECYT-PRONACES/61533/2020. The authors thank the referee for useful comments and suggestions that helped to improve this paper.


\bibliography{campos_electricos_bib}
\bibliographystyle{unsrt}   
\end{document}